\documentclass[twoside]{article}

\evensidemargin
\oddsidemargin
\advance \baselineskip 0pt plus 0.1pt minus 0.1pt
\advance \baselineskip 0pt plus 0.1pt minus 0.1pt
\input epsf
\usepackage[T2A]{fontenc}
\usepackage[cp1251]{inputenc}
\usepackage[english]{babel}

\usepackage{graphicx}
\usepackage{amsmath}
\usepackage{latexsym}
\usepackage{amsbsy}
\usepackage{amsfonts}
\usepackage{amssymb}
\usepackage[mathcal]{eucal}

\date{}

\begin{document}
\def\Z{\hbox{{\sf Z}\kern-0.4em {\sf Z}}}

\title{A chain of strongly correlated $SU(2)_{4}$ anyons: \\
Hamiltonian and Hilbert space of states}
\author{L. Martina, A. Protogenov\footnote{email: alprot@appl.sci-nnov.ru}, V. Verbus  \\
\\
{\fontsize{10pt}{12pt}\selectfont
{\it Dipartimento di Fisica, Universita del Salento, Lecce, Italy
}\/}\\
{\fontsize{10pt}{12pt}\selectfont
{\it Institute of Applied Physics, RAS, Nizhny Novgorod, Russia
}\/}\\
{\fontsize{10pt}{12pt}\selectfont
{\it Institute for Physics of Microstructures, RAS, Nizhny Novgorod, Russia
}}\/
}
\maketitle
\begin{abstract}
One-dimensional lattice model of $SU(2)_{4}$ anyons containing a transition into
the topological ordered phase state is considered. An effective low-energy Hamiltonian
is found for half-integer and integer indices of the type of strongly correlated non-Abelian
anyons.  The Hilbert state space properties in the considered modular tensor category are studied.
\\ [5pt] PACS: 05.30.Pr, 11.25.Hf, 05.50.+q, 03.65.Vf, 64.60.De.

\end{abstract}

\section{Introduction}

In basic models of quantum magnets, the states of particles with half-integer and integer
values of spin, which are distributed over lattice sites, correspond to the fundamental or adjoint
irreducible representation of the $SU(2)$ group and are classified according to odd or even
representations of the permutation group. In spatially one-dimensional or two-dimensional systems,
instead of the permutation group, we deal with the braid group and its Abelian and non-Abelian irreducible
representations. Therefore, in the generalized Heisenberg or Ising models, neither fermions, nor bosons
are distributed over lattice sites, but the quantum states of particles realizing irreducible
representations of the braid group and corresponding to fractional statistics.

The set of quantum numbers indicating the type of these so-called anyon states comprises
integers specifying the structure of quantum groups, e.g., $SU(2)_{k}$. Here, $k$ is the
$SU(2)_{k}$ level of the Wess-Zumino-Witten-Novikov theory. This number which coincides with
the coefficient in the Chern-Simons action, determines the braiding degree of excitation
worldlines in the $(2+1)D$ systems. When $k=1$, fermion worldlines are parallel. If $k=2$,
worldlines are linked twice and we deal with semi-fermion quasiparticles, the so-called
semions. Such anyons are primary fields of the $SU(2)_{2}$ conformal field
theory as well as that of the transverse field Ising model at the critical point \cite{NSSFS,RSW,Sachdev}.

Anyon states with $k=3$ form the Fibonacci family. Detailed analysis of these non-Abelian
states \cite{NSSFS,Pres} and generalization of the Heisenberg Hamiltonian to the case of interacting Fibonacci
anyons are presented in Refs. \cite{FTLTKWF,TTZL}.
Paper \cite{GATLTW} is devoted to collective states for a higher level of $k$
in the model of interacting $SU(2)_{5}$ anyons belonging to the sector with integer spins.
A characteristic feature of anyon systems for $k>2$ is the presence of a term describing
the three-body interaction in the Hamiltonian. Another important property is the fact that the
squared total quantum dimension of states with $k=1,2,4$ is an integer.
In this case, the density of braid group representations is such that it does not provide universal
quantum calculations \cite{FLW1,FLW2}. For $k \to \infty $, the Heisenberg Hamiltonian is restored for particles with the spin $1/2$, while for particles with an arbitrary spin, we have its polynomial variant \cite{Faddeev}.
To elucidate the Hamiltonian structure of the theory for intermediate values of $k$, in this paper we focus on the theory of anyons in the case of $k=4$.

Studying of anyon systems for small $k$ is important for several reasons. First of all, due to their
possible use for decoherence-free quantum computations based on the rigidity of the wave-function
phase of non-Abelian excitations \cite{Kitaev1}.
For this reason, anyons of the transverse field Ising model and
the $SU(2)_{k}$ theory with $k=2$ differing by the sign of the Frobenius-Schur indicator and
Fibonacci anyons for $k=3$ were analyzed \cite{TAFHLT} in detail as probable candidates for the role of non-Abelian states in the fractional quantum Hall effect with the filling factor $\nu = 2 + \frac{k}{k+2}$.
Phase effects in interferometry \cite{BSS,BFN} and tunneling \cite{Bonderson}
of non-Abelian anyon states for finding experimental consequences were studied
in a number of recent papers (see also review \cite{NSSFS}).
Discussion of the criteria of anyon deconfinement \cite{BS}, search of correspondences between the lattice
theory at the critical point and the conformal field theory determined on coset-spaces as the source
of the nucleation effects in parent two-dimensional spin liquid and edge modes at their
boundaries \cite{GATLTW,BSH}, as well as detailed description of topology driven quantum phase
transitions \cite{GTKLTW} in time-reversal invariant systems (in quantum doubles) are subjects
of recent papers in this challenging field.

In this paper, which employs the approach developed in Refs. \cite{TTZL,GATLTW},
we have found the Hamiltonian and the Hilbert space of states in the $SU(2)_{4}$ theory
for chains of half-integer and integer spins indicating sectors
of interacting anyons. Since we are interested in the low-energy
states, the effective low-energy Hamiltonian has the form of projectors on the
ground state with a fixed spin. Some aspects of the problem associated with representation of the
Hamiltonian by means of projectors of the Temperley-Lieb algebra have been discussed
in Refs. \cite{FTLTKWF,GTKLTW}. Maps of Hamiltonians \cite{Werner,Kivelson} of the exact solvable quantum
models \cite{Kitaev1,Kitaev2}, study of correlation functions \cite{Baskaran},
and classification of the toloplogical order \cite{Wen} and
quantum phase transitions are also considered.

We will pay attention to some universal features of externally different models and discuss a
correspondence to the conformal field theory in description of the braiding effects.
The main employed tools belong to the theory of modular tensor categories, in terms of which the anyon
theory is formulated now. Therefore, in the next section we present some technique from the theory of
modular tensor categories for describing the Hamiltonian and the eigenstate space considered in the third and
the fourth sections. In conclusion, we discuss the unsolved problems related to the universal representation of
Hamiltonian dynamics of the exact integrable spin models describing the topological ordered phase states,
which are characterized by non-Abelian anyon excitations.

\section{Chain of $SU(2)_{4}$-anyons}

Let us consider the Klein-type Hamiltinian family  \cite{Klein}

\begin{equation}
\label{eq1}
H = -\sum_{i=1;j}^{N}g_{j}P_{i}^{(j)} \, .
\end{equation}
Here, index $i$ enumerates links of the chain, $\{g_{j}\}$ are the coupling constants, and $P_{i}^{(j)}$
are projectors on states with the spin $j$.

In systems with non-Abelian quasiparticles, strongly correlated states form some part of the low-energy
Hilbert space. Since we are primarily interested in the low-energy limit, this property of
non-Abelian anyons is an argument for representation of effective Hamiltonian (\ref{eq1}) by means of operators
with two eigenvalues corresponding to the ground and excitated states. Besides, when the considered
particles are spatially separated, the low-energy space proves to be degenerate; constrains are
imposed on the degree of degeneracy, which are characterized by topological quantum numbers.

One of the methods for describing the topology effects related to braiding of quasiparticle
worldlines is based on the use of Temperley-Lieb algebra operators. The Temperley-Lib algebra
operators $e_{i}$ satisfy the commutation relations \cite{Temperley}

$$e_{i}^{2}=de_{i} \, ,$$ $$e_{i}e_{i+1}e_{i}=e_{i} \, \, \, \, ,
e_{i}e_{j}=e_{j}e_{i}, \, \, (j \neq i \pm 1) \, .$$
This implies that the operators $P_{i}=e_{i}/d$ are projectors: $P_{i}^{2}=P_{i}$. Here, the
isotopic parameter is $d=2\cos (\frac{\pi}{k+2})$.

The Hamiltonian for a small anisotropy parameter $x$ can be obtained using the transfer matrix
$T=\tau_{1}\tau_{3} \cdots \tau_{2N-1}I_{0} \tau_{2}\tau_{4} \cdots \tau_{2N-2}I_{2N}$
with $\tau_{i}=I_{i}+xe_{i}$ and the identity matrix $I_{i}$.
The Temperley-Lieb algebra operators $e_{i}$ are defined here \cite{Jones,Pasq,Kuniba} as

\begin{eqnarray}
\nonumber
{\bf e}[i] | j_{i-1} j_i j_{i+1} \rangle
&=&
\sum_{{j'}_i}
\left({\bf e}[i]^{j_{i+1}}_{j_{i-1}}\right)_{j_i}^{{j'}_i}
  |j_{i-1} {j'}_i j_{i+1} \rangle
\\
\quad
\left({\bf e}[i]^{j_{i+1}}_{j_{i-1}}\right)_{j_i}^{{j'}_i}
&=&
\delta_{j_{i-1},j_{i+1}}
\ \
\sqrt{
\frac{S^0_{j_i} S^0_{{j'}_i}}
{S^0_{j_{i-1}} S^0_{{j}_{i+1}}} \, ,
}
\label{eq2}
\end{eqnarray}
where $S^{j'}_j = \sqrt{\frac{2}{k+2}} \
\sin \left(\pi \frac{(2j+1)(2{j'}+1)}{k+2}\right)$ is the modular $S$-matrix $SU(2)_{k}$ of
the conformal field theory.

It follows from expressions (\ref{eq2}) that in the considered weight representation of the operators
$e_{i}$, anyon degrees of freedom are distributed over the three neighboring links $(i-1,i,i+1)$ of
the chain. Their filling by some value of the spin $a$ from the anyon-type set $j=0,1/2,1, ...,k/2$ will
be used below as the collective label of the quantum states $|j_{i-1},j_{i},j_{i+1}>$. The relationship
between the admissible values from the set $\{a,b,c \}$ of the angular momentum values is
determined by the fusion rules (tensor product) of the two anyon states in the $SU(2)_{k}$
theory:

\begin{equation}
\label{eq3}
j_{1}\times j_{2}=|j_{1}-j_{2}|+(|j_{1}-j_{2}|+1)+\ldots + {\rm min}(j_{1}+j_{2},k-j_{1}-j_{2}) \, .
\end{equation}
Composition of the fields truncated by the level $k$ after fusion is such
that instead of the standard result in the case of the $SU(2)$ tensor product, we have a smaller number
of fields at the output.

The fusion rules $a \times b = \sum_{c}N^{c}_{ab}c$ of primary fields for possible values
of the index $j$ for $k=4$, the quantum dimensions $d_{j}=[2j+1]_{q}=
\frac{q^{(j+1/2)}-q^{-(j+1/2)}}{q^{1/2}-q^{-1/2}}=
\sin [ \frac {\pi (2j+1)}{k+2} ]/  \sin [ \frac {\pi} {k+2} ]$ (for $q=e^{2\pi i/(k+2)}$)
being the largest eigenvalues of the matrix $(N_{a})^{c}_{b}$ with the total quantum dimension $\mathcal{D} = \sqrt{\sum_{a} d_{a}^{2}}
= \frac{\sqrt{\frac{k+2}{2}}}{\sin \left( \frac{\pi}{k+2} \right)}=2\sqrt{3}$, and the conformal
scale dimensions $h_{j}=\frac{j(j+1)}{k+2}$ determining the topological spin
$\theta_{j}=e^{2\pi i h_{j}}$ are summarized in the table

\begin{table}[hbt]
\begin{center}
$\begin{array}{|l|l}
\hline
\begin{array}{l|ll}
j=0&d_0=1&h_0=0  \\
j=1/2&d_{1/2}=\sqrt{3}&h_{1/2}=\frac{1}{8}\\
j=1&d_1=2&h_1=\frac{1}{3}\\
j=3/2&d_{3/2}=\sqrt{3}&h_{3/2}=\frac{5}{8}\\
j=2&d_2=1&h_2=1
\end{array}\\
\hline
\hline
\begin{array}{llll}
1/2\times 0=1/2~&& \\
1/2\times 1/2=0+1~& 1\times 0=1~& \\
1/2\times 1=1/2+3/2&1\times 1=0+1+2~&3/2\times 0=3/2~& \\
1/2\times 3/2=1+2&1\times 3/2=1/2+3/2&3/2\times 3/2=0+1&2\times 0=2 \\
1/2\times 2=3/2&1\times 2=1&3/2\times 2=1/2 & 2\times 2=0 \\
\end{array}\\
\hline
\end{array}$
\end{center}
\caption{\footnotesize Anyon types numbered by values of the index $j$, quantum and conformal dimensions,
and fusion rules for the $SU(2)_4$ theory.}
\label{table}
\end{table}

When two fields fuse, the vector space $V_{ab}^{c}$ of the dimension
${\rm dim} V_{ab}^{c}$ equal to $N^{c}_{ab}$ occurs. In our case, $N^{c}_{ab}=0,1$. For the
non-Abelian states indexed in the table by the spin values $j$, $d_{j} > 1$. Non-Abelian anyons
are also characterized by a multiplicity of allowed channels of fusion (splitting)
($\sum_{c}N^{c}_{ab} \ge 2$ for some value of the spin $b$).

Splitting of the anyon $b$ into anyon states with the spins $a,b,c$ corresponds to the space
$V_{d}^{abc}$, which can be represented in the form of tensor products of two splitted anyon spaces
by matching appropriate pairs of indices. This can be done by two isomorphic ways:

\begin{equation}
\label{eq4}
V_{d}^{abc}\cong \bigoplus\limits_{e}V_{e}^{ab}\otimes V_{d}^{ec}\cong
\bigoplus\limits_{f}V_{d}^{af}\otimes V_{f}^{bc},
\end{equation}

To introduce the notion of associativity at the level of splitting spaces, it is necessary to
describe the set of unitary isomorphisms between different decompositions, which can be considered
as a change of the basis. These isomorphic transformations can be written as a diagram 
using the so-called $F$-matrix:  

\begin{equation}
\label{eq5}
\end{equation}

\begin{figure}[h]
  \center{
      \includegraphics[width=0.35\columnwidth]{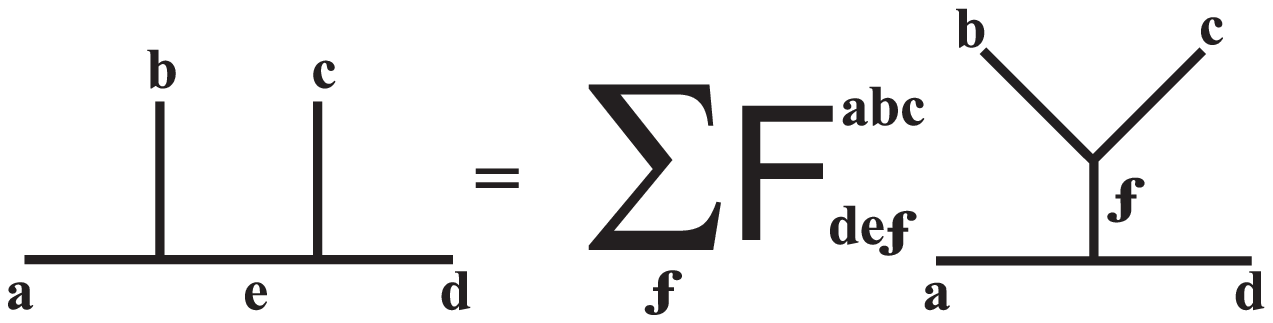}
  }
\end{figure}

The $F$-matrix satisfies the pentagonal equation

\begin{equation}
\label{eq6}
\sum_{n}F^{mlq}_{kpn}F^{jip}_{mns}F^{jsn}_{lkr}=F^{jip}_{qkr}F^{riq}_{mls} \, \, \, ,
\end{equation}
which has the solution \cite{KirResh} proportional to the $q$-deformed analog of the $6j$-symbols:

\begin{equation}
\label{eq7}
F_{j \, j_{12} \,j_{23}}^{j_{1} \, j_{2} \, j_{3}}
=\left( -1\right) ^{j_{1}+j_{2}+j_{3}+j}\sqrt{\left[
2j_{12}+1\right] _{q}\left[ 2j_{23}+1\right] _{q}}\left\{
\begin{array}{ccc}
j_{1} & j_{2} & j_{12} \\
j_{3} & j & j_{23}%
\end{array}
\right\} _{q} \, ,
\end{equation}

$$
\left\{
\begin{array}{ccc}
j_{1} & j_{2} & j_{12} \\
j_{3} & j & j_{23}%
\end{array}%
\right\} _{q}=\Delta \left( j_{1},j_{2},j_{12}\right) \Delta \left(
j_{12},j_{3},j\right) \Delta \left( j_{2},j_{3},j_{23}\right) \Delta \left(
j_{1},j_{23},j\right)
$$

$$
\times
\sum\limits_{n}\left\{ \frac{\left( -1\right) ^{n}\left[ n+1\right] _{q}!}{%
\left[ n-j_{1}-j_{2}-j_{12}\right] _{q}!\left[ n-j_{12}-j_{3}-j\right] _{q}!%
\left[ n-j_{2}-j_{3}-j_{23}\right] _{q}!\left[ n-j_{1}-j_{23}-j\right] _{q}!}%
\right.
$$

$$
\times \left. \frac{1}{\left[ j_{1}+j_{2}+j_{3}+j-n\right] _{q}!%
\left[ j_{1}+j_{12}+j_{3}+j_{23}-n\right] _{q}!\left[ j_{2}+j_{12}+j+j_{23}-n%
\right] _{q}!}\right\}_{\phantom{g}} ,
$$

$$
\Delta \left( j_{1},j_{2},j_{3}\right) =\sqrt{\frac{%
\left[ -j_{1}+j_{2}+j_{3}\right] _{q}!\left[ j_{1}-j_{2}+j_{3}\right] _{q}!%
\left[ j_{1}+j_{2}-j_{3}\right] _{q}!}{\left[ j_{1}+j_{2}+j_{3}+1\right]
_{q}!}}^{\phantom{T}}_{\phantom{g}},\quad \quad \quad \left[ n\right] _{q}! \equiv \prod\limits_{m=1}^{n}%
\left[ m\right] _{q} \, .
$$

In the following two sections, we will calculate the $F$-matrix different from zero or unity
by using Eq. (\ref {eq7}). We have not presented here the expressions for the braiding $R_{ab}^{c}$-matrix
of two anyons and the hexagon equation, because the latter
is a consequence \cite{Furusho} of the pentagon equation.

\section{The Hamiltonian and the Hilbert space of states}

When constructing effective Hamiltonian (\ref{eq1}), we employ
the representation \cite{FTLTKWF}

\begin{equation}
\label{eq8}
\langle a^{\prime},f^{\prime},d^{\prime}|P^{(j)}|a,e,d \rangle = F^{a^{\prime}bc}_{d^{\prime}jf^{\prime}}
F^{abc}_{dej} \delta_{a,a^{\prime}}  \delta_{d,d^{\prime}}
\end{equation}
for projectors in the channel with the total spin $j$, using the $F$-matrix. In Eq. (\ref{eq8}),
the summation over repetitive indices is absent. In the case of the Heisenberg Hamiltonian
$H=\sum_{i}({\bf S}_{i}{\bf S}_{i+1})$ for the spin equal to unity, we have
$({\bf S}_{i}{\bf S}_{i+1})=3P_{i}^{(2)}+P_{i}^{(1)}-2$. For the Affleck-Kennedy-Lieb-Tasaki model
$H=\sum_{i}P_{i}^{(2)}$. In our case, the superscripts $(b,c)$ in (\ref{eq8}) will acquire the value
$1/2$. In the sector with an integer value of the indices,
these labels of the anyon species will be as follows: $b=c=1$.

Let us consider first the half-integer values $b=c=1/2$ of the anyon index in the $F$-matrix. In this
case, the admissible state space $\{|a,f,d\rangle \}$ is fourteen-dimensional and represents the
following state set shown in the right-hand side of Eq. (\ref{eq5}):

\begin{equation}
\label{eq9}
\{|0,0,0\rangle, |0,1,1\rangle, |1/2,0,1/2\rangle,|1/2,1,1/2\rangle,|1/2,1,3/2\rangle,|1,1,0\rangle,|1,0,1\rangle,
\end{equation}
$$
|1,1,1\rangle, |1,1,2\rangle, |3/2,1,1/2\rangle,|3/2,0,3/2\rangle,|3/2,1,3/2\rangle,|2,1,1\rangle,|2,0,2\rangle\} \, .
$$
The Hilbert space $\{|a,e,d\rangle \}$ belonging to the left-hand side of Eq. (\ref{eq5})
consists of the following states:

\begin{equation}
\label{eq10}
\{|0,1/2,0\rangle, |0,1/2,1\rangle, |1/2,0,1/2\rangle,|1/2,1,1/2\rangle,|1/2,1,3/2\rangle,|1,1/2,0\rangle,|1,1/2,1\rangle,
\end{equation}
$$
|1,3/2,1\rangle, |1,3/2,2\rangle, |3/2,1,1/2\rangle,|3/2,1,3/2\rangle,|3/2,2,3/2\rangle,|2,3/2,1\rangle,|2,3/2,2\rangle\} \, .
$$

Let us write out the $F$-matrix components equal to unity:

\begin{equation}
\label{eq11}
F^{0\frac{1}{2}\frac{1}{2}}_{0\frac{1}{2}0}=
F^{0\frac{1}{2}\frac{1}{2}}_{1\frac{1}{2}1}=F^{\frac{1}{2}\frac{1}{2}\frac{1}{2}}_{\frac{3}{2}11}=
F^{1\frac{1}{2}\frac{1}{2}}_{0\frac{1}{2}1}=
\end{equation}
$$
F^{1\frac{1}{2}\frac{1}{2}}_{2\frac{3}{2}1}=
F^{\frac{3}{2}\frac{1}{2}\frac{1}{2}}_{\frac{1}{2}11}=
F^{2\frac{1}{2}\frac{1}{2}}_{1\frac{3}{2}1}=
F^{2\frac{1}{2}\frac{1}{2}}_{2\frac{3}{2}0} =1 \, .
$$

In the case of $k=4$ and $b=c=1/2$, a nontrivial contribution to the $F$-matrix structure is
made by three two-dimensional matrices $M_{j}$:

\begin{equation}
\label{eq12}
M_{\frac{1}{2}}=F^{\frac{1}{2}\frac{1}{2}\frac{1}{2}}_{\frac{1}{2}ef} =
\left(\begin{array}{cc}
F^{\frac{1}{2}}_{\frac{1}{2}00} & F^{\frac{1}{2}}_{\frac{1}{2}01} \\
F^{\frac{1}{2}}_{\frac{1}{2}10} & F^{\frac{1}{2}}_{\frac{1}{2}11} \\
\end{array}
\right) =
\frac{1}{\sqrt{3}}\left(\begin{array}{cc}
-1 & \sqrt{2} \\
\sqrt{2} & 1 \\
\end{array}
\right) \, ,
\end{equation}

\begin{equation}
\label{eq13}
M_{1}=F^{1\frac{1}{2}\frac{1}{2}}_{1ef} =
\left(\begin{array}{cc}
F^{1}_{1\frac{1}{2}0} & F^{1}_{1\frac{1}{2}1} \\
F^{1}_{1\frac{3}{2}0} & F^{1}_{1\frac{3}{2}1} \\
\end{array}
\right) =
\frac{1}{\sqrt{2}}\left(\begin{array}{cc}
-1 & 1 \\
1 & 1 \\
\end{array}
\right) \, ,
\end{equation}

\begin{equation}
\label{eq14}
M_{\frac{3}{2}}=F^{\frac{3}{2}\frac{1}{2}\frac{1}{2}}_{\frac{3}{2}ef} =
\left(\begin{array}{cc}
F^{\frac{3}{2}}_{\frac{3}{2}10} & F^{\frac{3}{2}}_{\frac{3}{2}11} \\
F^{\frac{3}{2}}_{\frac{3}{2}20} & F^{\frac{3}{2}}_{\frac{3}{2}21} \\
\end{array}
\right) =
\sqrt{\frac{2}{3}}\left(\begin{array}{cc}
-1 & \frac{3^{1/4}}{2} \\
\frac{3^{1/4}}{2} & 1 \\
\end{array}
\right) \, .
\end{equation}
For simplicity of the notation, in the three latter formulas we have omitted fixed indices of the elements
of the $F$-matrices. In calculation of the elements of the matrices in these formulas, we used explicit
expression (\ref{eq7}) for the $q$-deformed values of the $6j$-symbols.

Let us employ the given values of the $F$-matrix to calculate the projectors by means of the $F$-matrix.
In the case of $b=c=1/2$ and the level $k=4$, the following two projectors prove to be nonzero:

\begin{equation}
\label{eq15}
P^{(0)}={\rm diag}\,\left(0,0,A,0,0,0,0,0,0,0,0,0,0\right) \, ,
\end{equation}

\begin{equation}
\label{eq16}
P^{(1)}={\rm diag}\,\left(0,0,B,1,0,0,0,0,1,C,0,0\right) \, ,
\end{equation}
where the $2 \times 2$ matrices $A,B,C$ are present in diagonals and have the forms

\begin{equation}
\label{eq17}
A=\left(\begin{array}{cc}
\left(F^{\frac{1}{2}}_{\frac{1}{2}00}\right)^{2} & F^{\frac{1}{2}}_{\frac{1}{2}00}F^{\frac{1}{2}}_{\frac{1}{2}01} \\
F^{\frac{1}{2}}_{\frac{1}{2}10}F^{\frac{1}{2}}_{\frac{1}{2}00} & F^{\frac{1}{2}}_{\frac{1}{2}01}F^{\frac{1}{2}}_{\frac{1}{2}10} \\
\end{array}
\right) =
\frac{1}{3}
\left(\begin{array}{cc}
1 & -\sqrt{2} \\
-\sqrt{2} & 2 \\
\end{array}
\right) \, ,
\end{equation}

\begin{equation}
\label{eq18}
B=\left(\begin{array}{cc}
F^{\frac{1}{2}}_{\frac{1}{2}10}F^{\frac{1}{2}}_{\frac{1}{2}01} & F^{\frac{1}{2}}_{\frac{1}{2}10}F^{\frac{1}{2}}_{\frac{1}{2}11} \\
F^{\frac{1}{2}}_{\frac{1}{2}11}F^{\frac{1}{2}}_{\frac{1}{2}10} & \left(F^{\frac{1}{2}}_{\frac{1}{2}11}\right)^{2} \\
\end{array}
\right) =
\frac{1}{3}\left(\begin{array}{cc}
2 & \sqrt{2} \\
\sqrt{2} & 1 \\
\end{array}
\right) \, ,
\end{equation}

\begin{equation}
\label{eq19}
C=\left(\begin{array}{cc}
F^{\frac{3}{2}}_{\frac{3}{2}10}F^{\frac{3}{2}}_{\frac{3}{2}11} & F^{\frac{3}{2}}_{\frac{3}{2}10}F^{\frac{3}{2}}_{\frac{3}{2}21} \\
\left(F^{\frac{3}{2}}_{\frac{3}{2}11}\right)^{2}
 & F^{\frac{3}{2}}_{\frac{3}{2}11}F^{\frac{3}{2}}_{\frac{3}{2}21} \\
\end{array}
\right) =
\frac{1}{3}\left(\begin{array}{cc}
-\sqrt{2} &  -2 \\
1 & \sqrt{2}  \\
\end{array}
\right) \, .
\end{equation}
The enumeration in Eqs. (\ref{eq15}), (\ref{eq16}) corresponds to listing of states in
Eq. (\ref{eq10}). As a result, the Hamiltonian of the chain of the $SU(2)_{4}$ anyons is as follows:

\begin{equation}
\label{eq20}
H = -\sum_{i=1}^{N}\left(g_{0}P_{i}^{(0)}+ g_{1}P_{i}^{(1)}\right)
\end{equation}
with projectors from Eqs. (\ref{eq15}), (\ref{eq16}).

\section{Integer anyon index}

In this section, we shall consider integer values of the index $c=b=1$ of anyon types. In this case,
the admissible state space $\{|a,f,d\rangle \}$ is nineteen-dimensional and represents the state set
shown in the right-hand side of Eq. (\ref{eq5}):

\begin{equation}
\label{eq21}
\{|0,0,0\rangle,
|0,1,1\rangle,
|0,2,2\rangle,
|1/2,0,1/2\rangle,
|1/2,1,1/2\rangle,
|1/2,1,3/2\rangle,
\end{equation}
$$
|1/2,2,3/2\rangle,
|1,0,1\rangle,
|1,1,1\rangle,
|1,2,1\rangle,
|1,1,0\rangle,
|1,1,2\rangle,
$$
$$
|3/2,1,1/2\rangle,
|3/2,2,1/2\rangle,
|3/2,0,3/2\rangle,
|3/2,1,3/2\rangle,
$$
$$
|2,2,0\rangle,
|2,1,1\rangle,
|2,0,2\rangle \} \, .
$$

The Hilbert state space $\{|a,e,d\rangle \}$ shown in the left-hand side of Eq. (\ref{eq5})
consists of the following states:

\begin{equation}
\label{eq22}
\{|0,1,0\rangle,
|0,1,1\rangle,
|0,1,2\rangle,
|1/2,1/2,1/2\rangle,
|1/2,3/2,1/2\rangle,
|1/2,1/2,3/2\rangle,
\end{equation}
$$
|1/2,3/2,3/2\rangle,
|1,0,1\rangle,
|1,1,1\rangle,
|1,2,1\rangle,
|1,1,0\rangle,
|1,1,2\rangle,
$$
$$
|3/2,1/2,1/2\rangle,
|3/2,3/2,1/2\rangle,
|3/2,1/2,3/2\rangle,
|3/2,3/2,3/2\rangle,
$$
$$
|2,1,0\rangle,
|2,1,1\rangle,
|2,1,2\rangle \} \, .
$$

The one-dimensional $F$-matrices equal to unity correspond to the following sequence of indices:

\begin{equation}
\label{eq23}
F^{011}_{010}=
F^{0\frac{1}{2}\frac{1}{2}}_{1\frac{1}{2}1}=
F^{\frac{1}{2}\frac{1}{2}\frac{1}{2}}_{\frac{3}{2}11}=
F^{1\frac{1}{2}\frac{1}{2}}_{0\frac{1}{2}1}=
\end{equation}
$$
F^{1\frac{1}{2}\frac{1}{2}}_{2\frac{3}{2}1}=
F^{\frac{3}{2}\frac{1}{2}\frac{1}{2}}_{\frac{1}{2}11}=
F^{2\frac{1}{2}\frac{1}{2}}_{1\frac{3}{2}1}=
F^{2\frac{1}{2}\frac{1}{2}}_{2\frac{3}{2}0} =1 \, .
$$

The four two-dimensional $F$-matrices ${\hat M}_{j}$ are equal to

\begin{equation}
\label{eq24}
{\hat M}_{\frac{1}{2}}=F^{\frac{1}{2}11}_{\frac{1}{2}ef} =
\left(\begin{array}{cc}
F^{\frac{1}{2}}_{\frac{1}{2}\frac{1}{2}0} & F^{\frac{1}{2}}_{\frac{1}{2}\frac{1}{2}1} \\
F^{\frac{1}{2}}_{\frac{1}{2}\frac{3}{2}0} & F^{\frac{1}{2}}_{\frac{1}{2}\frac{3}{2}1} \\
\end{array}
\right) =
\frac{1}{\sqrt{2}}\left(\begin{array}{cc}
-1 & 1 \\
1 & 1 \\
\end{array}
\right) \, ,
\end{equation}

\begin{equation}
\label{eq25}
{\hat M}_{\frac{1}{2}}^{'}=F^{\frac{1}{2}11}_{\frac{3}{2}ef} =
\left(\begin{array}{cc}
F^{\frac{1}{2}}_{\frac{3}{2}\frac{1}{2}1} & F^{\frac{1}{2}}_{\frac{3}{2}\frac{1}{2}2} \\
F^{\frac{1}{2}}_{\frac{3}{2}\frac{3}{2}1} & F^{\frac{1}{2}}_{\frac{3}{2}\frac{3}{2}2} \\
\end{array}
\right) =
\frac{1}{\sqrt{2}}\left(\begin{array}{cc}
-1 & 1 \\
1 & 1 \\
\end{array}
\right) \, ,
\end{equation}

\begin{equation}
\label{eq26}
{\hat M}_{\frac{3}{2}}=F^{\frac{3}{2}11}_{\frac{1}{2}ef} =
\left(\begin{array}{cc}
F^{\frac{3}{2}}_{\frac{1}{2}\frac{1}{2}1} & F^{\frac{3}{2}}_{\frac{1}{2}\frac{1}{2}2} \\
F^{\frac{3}{2}}_{\frac{1}{2}\frac{3}{2}1} & F^{\frac{3}{2}}_{\frac{1}{2}\frac{3}{2}2} \\
\end{array}
\right) =
\frac{1}{\sqrt{2}}\left(\begin{array}{cc}
-1 & 1 \\
1 & 1 \\
\end{array}
\right) \, ,
\end{equation}

\begin{equation}
\label{eq27}
{\hat M}_{\frac{3}{2}}^{'}=F^{\frac{3}{2}11}_{\frac{3}{2}ef} =
\left(\begin{array}{cc}
F^{\frac{3}{2}}_{\frac{3}{2}\frac{1}{2}0} & F^{\frac{3}{2}}_{\frac{3}{2}\frac{1}{2}1} \\
F^{\frac{3}{2}}_{\frac{3}{2}\frac{3}{2}0} & F^{\frac{3}{2}}_{\frac{3}{2}\frac{3}{2}1} \\
\end{array}
\right) =
\frac{1}{\sqrt{2}}\left(\begin{array}{cc}
1 & -1 \\
-1 & -1 \\
\end{array}
\right) \, .
\end{equation}

The $F$-symbol also contains a three-dimensional matrix having the form

\begin{equation}
\label{eq28}
{\hat M}_{1}=F^{111}_{1ef}=
\left(\begin{array}{ccc}
F^{1}_{100} & F^{1}_{101} & F^{1}_{102} \\
F^{1}_{110} & F^{1}_{111} & F^{1}_{112} \\
F^{1}_{120} & F^{1}_{121} & F^{1}_{122} \\
\end{array}
\right)=
\frac{1}{2}\left(\begin{array}{ccc}
1 & -{\sqrt 2} & 1 \\
 -{\sqrt 2} & 0 & {\sqrt 2} \\
1 & {\sqrt 2} & 1 \\
\end{array}
\right) \, .
\end{equation}

The projectors ${\hat P}^{(j)}$ with $j=0,1,2$ and $e=0,1,2$, $f=0,1,2$ determined by Eq. (\ref{eq28})
in the three-dimensional subspace are equal to

\begin{equation}
\label{eq29}
{\hat P}^{(0)}=F^{111}_{1e0}F^{111}_{10f}=
\left(\begin{array}{ccc}
F_{00}F_{00} & F_{00}F_{01} & F_{00}F_{02}  \\
F_{10}F_{00} & F_{10}F_{01} & F_{10}F_{02}  \\
F_{20}F_{00} & F_{20}F_{01} & F_{20}F_{02}  \\
\end{array}
\right)=
\frac{1}{4}\left(\begin{array}{ccc}
1 & -{\sqrt 2} & 1   \\
-{\sqrt 2} & 2 & -{\sqrt 2}  \\
1 & -{\sqrt 2} & 1  \\
\end{array}
\right)
\, ,
\end{equation}

\begin{equation}
\label{eq30}
{\hat P}^{(1)}=F^{111}_{1e1}F^{111}_{11f}=
\left(\begin{array}{ccc}
F_{01}F_{10} & F_{01}F_{11} & F_{01}F_{12}  \\
F_{11}F_{10} & F_{11}F_{11} & F_{11}F_{12}  \\
F_{21}F_{10} & F_{21}F_{11} & F_{21}F_{12}  \\
\end{array}
\right)=
\frac{1}{2}\left(\begin{array}{ccc}
1 & 0 & -1   \\
0 & 0 & 0  \\
-1 & 0 & 1  \\
\end{array}
\right) \, ,
\end{equation}

\begin{equation}
\label{eq31}
{\hat P}^{(2)}=F^{111}_{1e2}F^{111}_{12f}=
\left(\begin{array}{ccc}
F_{02}F_{20} & F_{02}F_{21} & F_{02}F_{22}  \\
F_{12}F_{20} & F_{12}F_{21} & F_{12}F_{22}  \\
F_{22}F_{20} & F_{22}F_{21} & F_{22}F_{22}  \\
\end{array}
\right) =
\frac{1}{4}\left(\begin{array}{ccc}
1 & {\sqrt 2} & 1   \\
{\sqrt 2} & 2 & {\sqrt 2}  \\
1 & {\sqrt 2} & 1  \\
\end{array}
\right)
\, .
\end{equation}
The projectors ${\hat P}^{(j)}$ dependent on the matrix ${\hat M}_{j}$ in two-dimensional
subspaces with half-integer indices $a$ and $d$ are zero:
${\hat P}^{(1/2)}={\hat P}^{'(1/2)}={\hat P}^{(3/2)}={\hat P}^{'(3/2)}=0$.

As a result, for $k=4$ and $a=b=1$ we have ${\bar P}^{(j)}=0$ for half-integer values of the
anyon type and the following nonzero projectors in the integer-valued sector

\begin{equation}
\label{eq33}
{\bar P}^{(0)}={\rm diag}\,\left(0,0,0,0,0,0,0,{\hat P}^{(0)},0,0,0,0,0,0,0,0,0\right) \, ,
\end{equation}

\begin{equation}
\label{eq34}
{\bar P}^{(1)}={\rm diag}\,\left(0,1,0,0,0,0,0,{\hat P}^{(1)},1,1,0,0,0,0,0,1,0\right) \, ,
\end{equation}

\begin{equation}
\label{eq35}
{\bar P}^{(2)}={\rm diag}\,\left(0,0,0,0,0,0,0,{\hat P}^{(2)},0,0,0,0,0,0,0,0,0\right) \, .
\end{equation}
where the three-dimensional matrices ${\hat P}^{(j)}$ are located on diagonals.

All aforesaid means that the Hamiltonian of the chain of the $SU(2)_{4}$-anyons for
the ladder indices $b=c=1$ has the form

\begin{equation}
\label{eq36}
H = -\sum_{i=1}^{N}\left(g_{0}{\bar P}_{i}^{(0)}+g_{1}{\bar P}_{i}^{(1)}+
g_{2}{\bar P}_{i}^{(2)}\right)
\end{equation}
with projectors from Eqs. (\ref{eq33}), (\ref{eq34}), and (\ref{eq35}).

\section{Conclusion}

The features of the Hamiltonian dynamics in the case of the $SU(2)_{4}$ theory can be revealed by comparing it
with the $SU(2)_{3}$ theory \cite{FTLTKWF} of Fibonacci anyons and the $SU(2)_{5}$ theory \cite{TTZL}.
In the latter case for $b=c=1$, of interest is the absence into total projectors
(\ref{eq33})-(\ref{eq35}) of the contribution made by projectors in the two-dimensional subspace.
Incontrast with the Hamiltonian of Fibonacci anyons, the one-dimensional projector is absent in the Hamiltonian
of the $SU(2)_{4}$ theory, when the zero energy is assigned to the state with $j=1$.

For the coinciding coupling constants in the Hamiltonian (\ref{eq36}), we have the model describing the
critical state; in the continuous limit, it coincides with the rational conformal field theory with
the central charge $c=4/5$. For the opposite sign of the coupling constants, the considered case $k=4$ in the
continuous limit corresponds to the $Z_{4}$ theory of parafermions with the central charge $c=1$. If
the coupling constants do not coincide, we can analyze the type of gapped phase states by assigning the
zero energy to the phase state with the spin $j=1$ in (\ref{eq20}), i.e., by setting $g_{1}=0$ in
(\ref{eq20}) and $g_{0}=0$ with the spin $j=0$ in (\ref{eq36}). This can be done by changing the angle
$\theta$ for parametrization of the coupling constants $g_{1}=\sin \theta$, $g_{2}=\cos \theta$
in (\ref{eq36}) as it was done in Refs. \cite{GATLTW,TTZL}.
We plan to consider this problem in a separate paper. In this paper, we focus on the study of the Hamiltonian representation of exact integrable anyon systems in the form of projectors in a more general formulation.

When discussing weight representation (\ref{eq2}) of the Temperley-Lieb algebra, we have already
mentioned that the Hamiltonian in the extremely anisotropic case $x \to 0$ at the boundary of some
regimes of the exact integrable $RSOS$-model \cite{ABF,Fendley} can be obtained using the $R$-matrix.
This matrix satisfies the Yang-Baxter equation

\begin{equation}
\label{eq37}
R_{12}R_{13}R_{23}=R_{23}R_{13}R_{12}
\end{equation}
In this equation, we employed standard notations $R_{12}=R\otimes R\otimes 1$, etc. with the
$R$ matrix realizing self-mapping of the space $V\otimes V$\rotatebox[origin=c]{180}{$\hookrightarrow $}
in itself.

In its turn, the $R$-matrix of the models describing topological ordered phase states can be
constructed using the solution \cite{Kashaev}

\begin{equation}
\label{eq38}
R_{12,34}=\left(F_{14}^{t_{4}}\right)^{-1}F_{13}F_{24}^{t}\left(F_{23}^{-1}\right)^{t_{2}}
\end{equation}
of the pentagon equation

\begin{equation}
\label{eq39}
F_{12}F_{13}F_{23}=F_{23}F_{12}  \, .
\end{equation}
Here, $R_{12,34} \in W\otimes W$, where $W=V\otimes V^{\star}$,
and the indices $t$ and $t_{i}$ mean, respectively, complete and partial transpositions in the $i$-th
space. The indices of the $F$-matrices in Eqs. (\ref{eq38}) and (\ref{eq39}) belong to
the four {\it faces of tetrahedron}. In the general case, the $F$-matrix also depends on the spectral parameter
$x$ \cite{Kashaev,Turaev}, the small value of which yields the Hamiltonian. For example, the case
\cite{Turaev}
\begin{equation}
\label{eq40}
R(x)=[x-1]_{q} + [x]_{q}e_{1}
\end{equation}
yields the known answers \cite{GTKLTW,FF}.

Let us consider the $(2+1)D$ models and the tensor $T_{ijkl}$ which depends on the indices $(i,j,k,l)$
belonging  to the tetrahedron faces. Each of these indices is the collective label comprising variables from
the sets $(a,b,c,d,e,f)$, $(i_{s},j_{s},k_{s},l_{s})$ with $s=1,2,3,4$ of the indices belonging respectively
to edges and vertices of each tetrahedron face. Employing the $T_{ijkl}$-tensors, we can find the
partition sum \cite{GuWen,LevinNave}
\begin{equation}
\label{eq41}
Z=Tr\,\, e^{-\beta H}=\sum_{ijkl}\,T_{jfei}T_{hgjk}T_{qklr}T_{lits} \cdots = tTr\,\otimes_{i}T
\end{equation}
by computing the tensor trace $tTr\,\otimes_{i}T$.
To carry out the calulations, anzats for the $T_{ijkl}$-tensor can be written as

\begin{equation}
\label{eq42}
T_{[i][j][k][l]}=F^{abc}_{def}\cdot \lambda (i_{1},j_{1},k_{1}) \cdot \lambda (i_{2},j_{2},l_{2}) \cdot
\lambda (i_{3},k_{3},l_{3})\cdot \lambda (j_{4},k_{4},l_{4}) \,.
\end{equation}

Since we are usually interested in the system behavior at the distances much larger than the
lattice constant, now we consider also the final stage of the renormalization group flow, which does not
contain data about the distribution of the degrees of freedom at small distances. The fixed surface at this
final stage of the renormalization group flow in the terms of the
functions from Eq. (\ref{eq42}) corresponds to the condition that all functions of local variables
tend to be constant.
In other words, all the functions are $\lambda \to 1$ up to normalization. This
conjecture  means that topological universality of the behavior at large distances occurs in
the infrared limit due to special contribution to the partition sum , which
{\it depends only on the $F$-matrix}.
Practical implementation of the computation program of the result of such normalization group flow is
related to the solution of the problem of finding a convenient relation between the variables
belonging to tetrahedron faces and the variables belonging to tetrahedron edges and vertices.
We plan to consider this problem in future.

\section{Acknowledgement}

We are grateful to R.M. Kashaev and V.G. Turaev for useful discussions.
The authors are also grateful to INFN for partial support in the framework of the
LE41 project, to the Abdus Salam International Centre for Theoretical Physics in Trieste and
the Galileo Galiley Institute for Theoretical Physics in Florence for hospitality.
This work was supported in part by a grant of the E.I.N.S.T.E.IN consortium (L.M. and A.P.),
the Russian Foundation for Basic Research (Grants No. 09-01-92426 and 09-01-00268, V.V. and А.P.),
and the program "Basic problems of nonlinear dynamics" of the Presidium of the Russian Academy of Sciences.

\end{document}